\documentclass{article}

\usepackage[aux]{rerunfilecheck}
\usepackage{amsmath}
\usepackage{amssymb}
\usepackage{mathtools}

\usepackage[numbers,sort&compress]{natbib}
\usepackage[autostyle]{csquotes}
\usepackage{xfrac}
\usepackage{float}
\usepackage{graphicx}
\usepackage{color}
\usepackage{bbm} 
\usepackage{placeins} 
\usepackage{booktabs}
\usepackage{geometry}

\usepackage{hyperref}
\usepackage{parskip} 
\usepackage{tikz-feynman}
\usepackage{textpos}

\title{\bf Radiative neutrino masses and the Cohen--Kaplan--Nelson bound}
\author{Patrick Adolf$^a$\thanks{patrick.adolf@tu-dortmund.de} , Martin Hirsch$^b$\thanks{mahirsch@ific.uv.es} , Heinrich P\"as$^a$\thanks{heinrich.paes@tu-dortmund.de}\\
\smallskip
\\
{\it $^a$ Fakultät für Physik, Technische Universität Dortmund,}
\\
{\it 44221 Dortmund, Germany}
\\
{\it $^b$ Instituto de F\'isica Corpuscular (IFIC),
Universidad de Valencia-CSIC,}
\\
{\it E-46980 Valencia, Spain}
}

\begin{document}
\maketitle
 \begin{textblock*}{3cm}(13cm,-8cm)
     DO-TH 23/08
   \end{textblock*}

\begin{abstract}
Recently, an increasing interest in UV/IR mixing phenomena has drawn attention to the range of validity 
of standard quantum field theory. Here we explore the consequences of such a limited range of validity in the
context of radiative models for neutrino mass generation. We adopt an argument first published by 
Cohen, Kaplan and Nelson that gravity implies both UV and IR cutoffs, apply it to the loop integrals describing radiative corrections, 
and demonstrate that this effect has significant consequences for the parameter space of radiative neutrino mass models. 
\end{abstract}

\section{Introduction}
\begin{figure}[!t]
    \centering
    \includegraphics[width=0.9\linewidth]{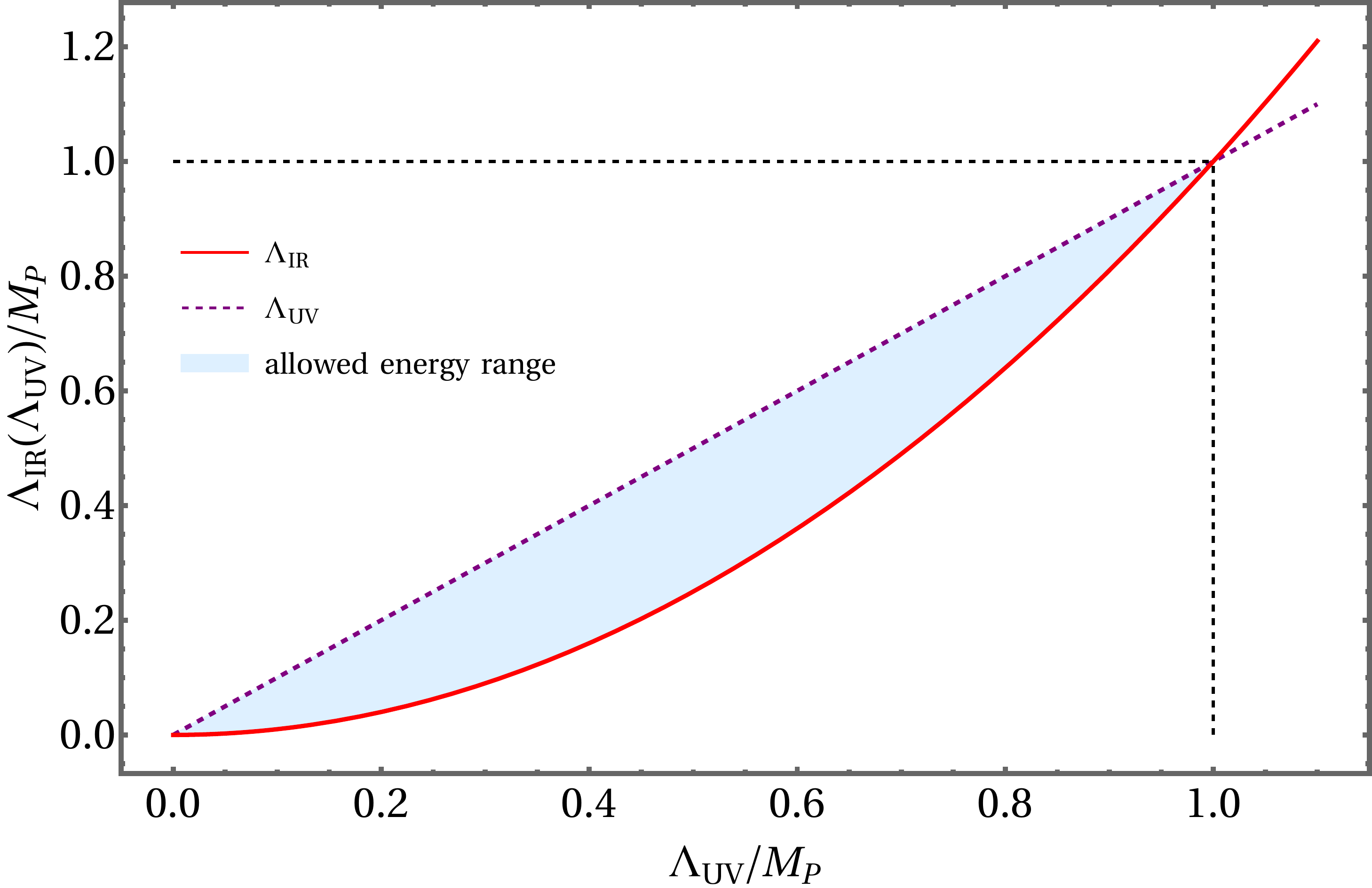}
    \caption{Allowed energy range accordingly to the CKN bound \eqref{eq:ckn}.}
    \label{fig:energy}
\end{figure}

In a seminal paper ~\cite{Cohen:1998zx} Cohen, Kaplan and Nelson (CKN) have argued that quantum field theories coupled to gravity feature not only a UV cutoff but also an IR cutoff, and that these cutoffs are related via the physics of black holes. As CKN argued, an IR cutoff follows from the holographic properties of black holes, according to which the maximum amount of information in a box of size $L$ does not increase
extensively with $\sim L^3$ but with the surface area $\sim L^2$. An even more stringent cutoff
results when one takes into account that quantum field theory (QFT) cannot describe black holes and 
adopts that the box size $L$ must always be larger than the Schwarzschild radius of the matter composition inside ~\cite{Cohen:1998zx},
\begin{equation}\label{eq:ckn}
    \Lambda_{I R} \geq \frac{\Lambda_{U V}^{2}}{M_{P}}\text,
\end{equation}
with $M_P$ being the Planck mass. This bound also ensures that
the entropy of the box is always lower than the black hole entropy and thus provides the allowed energy range in QFT shown in figure \ref{fig:energy}.

The CKN bound is interesting theoretically as one of a few phenomena that don't respect the usually assumed separation of scales in QFTs, an effect known as UV/IR mixing, but may also have interesting phenomenological consequences as the cutoff entailed affects the  
calculation of loop integrals by imposing momentum cutoffs~\cite{Cohen:2021zzr,Banks:2019arz}
\begin{equation}
\int_{0}^{\infty}dl\frac{l^3}{(l^2+\Delta)^3}\rightarrow\int_{\frac{1}{L_{IR}}}^{\Lambda_{UV}}dl\frac{l^3}{(l^2+\Delta)^3} \text.
\end{equation}
One notable example regards the quantum corrections to the dark energy density propelling the accelerated expansion of the universe that
are obtained in the correct order of magnitude $\rho_V \sim \left(10^{-3} \;\mathrm{eV}\right)^4$ when a IR cutoff corresponding to the cosmic horizon size is chosen. Another important application is the 1-loop correction to the magnetic moment of SM fermions, that instead of the 
well-known Schwinger term~\cite{Schwinger:1948iu} $\frac{\alpha}{2\pi}$, results as an expression depending on both cutoffs, 
\begin{equation}\label{eq:exact}
    \delta(g-2)_l=\frac{\alpha}{2\pi}\left(-\frac{\Lambda_{IR}}{m_l}\arctan\left(\frac{m_l}{\Lambda_{IR}}\right)+\frac{\Lambda_{UV}}{m_l}\arctan\left(\frac{m_l}{\Lambda_{UV}}\right)\right) \text.
\end{equation}
Note that the exact expression given in Eq.~\eqref{eq:exact} differs from the approximations for the limit $\Lambda_{UV} \rightarrow \infty$ and $\Lambda_{IR} \rightarrow 0$
adopted in~\cite{Cohen:1998zx,Cohen:2021zzr,Banks:2019arz}, while the numerical difference between this expressions is negligible for the energy range studied in this work. 

Since there is no theory that determines the exact choice of $\Lambda_{UV}$ (though there exist arguments to motivate a specific choice, see e.g. \cite{Kephart:2022vfr}), the two possible options are a universal $\Lambda_{UV}$ cutoff, which should be compatible with all measurements, or a process or particle dependent cutoff. To discuss the former possibility, we check the energy range for the UV cutoff in the case of the magnetic moment of electrons and muons (we do not consider the tau lepton, since the experimental measurements are not precise enough yet).
It is well-known that the experimentally measured value of $a_\mu$ differs from the SM prediction by more than 4 $\sigma$~\cite{Muong-2:2021ojo}. The magnetic moment from Eq.~\eqref{eq:exact} is always smaller than the Schwinger term and thus cannot explain this difference. Since the CKN correction is different from zero for any choice of the UV cutoff, we decided to estimate an allowed range for the UV cutoff, by minimizing the change of the Schwinger term. We then postulate that the SM prediction should not change by the inclusion of the CKN correction by more than twice the combined theoretical and experimental error.
For the electron we use the actual difference according to Refs.~\cite{Li:2021koa,Aoyama:2019ryr,Hanneke:2008tm} as the maximum possible range
\begin{align}
    \delta a_{e}&:=a_{e}^{\exp}-a_{e}^{\mathrm{SM}}(Cs)=(-8.8 \pm 3.6) \cdot 10^{-13} \label{eq:el}\\
    \delta a_{\mu}&:=a_{\mu}^{\exp }-a_{\mu}^{\mathrm{SM}}=(251\pm 59) \cdot 10^{-11}\text,\label{eq:mu}
\end{align} 
since the CKN bound could explain the difference. We have to mention, that the current situation of the magnetic moment of the electron is unclear, as there is a new measurement with 5 $\sigma$ deviation~\cite{Morel:2020dww}.
For definiteness, we apply~\eqref{eq:el} and \eqref{eq:mu} and obtain \(\Lambda_{UV} \sim (10^{1} - 10^{3})\;\mathrm{GeV}\) for the electron and 
\(\Lambda_{UV} \sim (10^{1} - 10^{6})\;\mathrm{GeV}\) for the muon. This rough estimate is sufficient for our purposes. If the choice of the UV cutoff is assumed to be independent of the process
considered (and if we don't assume any other new physics effects), the UV cutoff needs to be around \(\Lambda_{UV} \sim (10^{1} - 10^{3})\;\mathrm{GeV}\) in order to not disturb the SM predictions for the magnetic moments. The corresponding IR cutoff follows as \(\Lambda_{IR} \sim (10^{-18} - 10^{-14})\;\mathrm{GeV}\).

\section{Radiative neutrino models with CKN bound}
In the following we investigate the influence of the CKN bound on radiative neutrino models \cite{Cai:2017jrq}. As the CKN bound has an impact on the calculation of loop integrals, it can be expected that this effect is particularly large for models generating neutrino masses at 1-loop.
To develop an intuition how relevant the CKN bound is for this class of models, we study four representative models from the large variety 
of models existing~\cite{Arbelaez:2022ejo}. We motivate our choice by the analysis in Ref. \cite{Bonnet:2012kz} where it has been shown 
that all models generating neutrino masses at 1-loop order can be categorized into four different topologies of Feynman diagrams. We thus
study one model for each topology, and choose the most prominent representative, respectively.

\subsection{Scotogenic model (T-3)}
The scotogenic model~\cite{Ma:2006km} only requires to add a scalar doublet ($\eta^{+},\eta^{0}$) and singlet fermions $N_i$ to the SM.
Furthermore this model provides two possible candidates for dark matter, namely either the lightest fermion or bosonic mass eigenstate, hence it is also interesting for the connection between neutrino masses and dark matter. To explicitly avoid tree-level masses of the neutrinos a dark symmetry $\mathbb{Z}_2$ is considered, under which the new particles transform as odd and the SM particles as even. 
For simplicity, we consider only one scalar doublet, but the model can be extended to an arbitrary number of doublets and singlets~\cite{Escribano:2020iqq}.

The Yukawa Lagrangian for the new fermion singlets is given by 
\begin{equation}
    \mathcal{L}_{N,Yukawa}=h_{i j}\left(\nu_{i} \eta^{0}-l_{j} \eta^{+}\right) N_{j}+h.c.
\end{equation}
and together with the Majorana mass terms of the singlet fermions
\begin{equation}
    \frac{1}{2} M_{i} N_{i} N_{i}+h.c.
\end{equation}
and the scalar potential
\begin{equation}
    \begin{aligned} V & =m_{1}^{2} H^{\dagger} H+m_{2}^{2} \eta^{\dagger} \eta+\frac{1}{2} \lambda_{1}\left(H^{\dagger} H\right)^{2}+\frac{1}{2} \lambda_{2}\left(\eta^{\dagger} \eta\right)^{2}+\lambda_{3}\left(H^{\dagger} H\right)\left(\eta^{\dagger} \eta\right) \\ & +\lambda_{4}\left(H^{\dagger} \eta\right)\left(\eta^{\dagger} H\right)+\frac{1}{2} \lambda_{5}\left[\left(H^{\dagger} \eta\right)^{2}+h.c.\right]\text,\end{aligned}
\end{equation}
one obtains the 1-loop diagram in figure \ref{fig:scoto} as the lowest order contribution to neutrino masses.
\begin{figure}[!t]
    \centering
    \begin{tikzpicture}[x=3cm, y=3cm]
        \begin{feynman}
        \vertex (ghost) at (0,0);
        \vertex (a) at (0,0.5) {\( \nu_L \)};
        \vertex (b) at (1,0.5); 
        \vertex (c) at (2,0.5);
        \vertex (d) at (3,0.5){\( \nu_L \)};
        \vertex (h1) at (1,1.5){H};
        \vertex (h2) at (2,1.5){H};
        \vertex (four) at (1.5,1);
        \diagram* {
        (a)--[fermion](b) --[anti majorana, edge label=\(N\)] (c);
        (d)--[fermion] (c);
        (four)--[scalar, quarter right, edge label'=\(\eta\)](b);
        (four)--[scalar, quarter left,edge label=\(\eta\)](c);
        (h1)--[scalar](four);
        (h2)--[scalar](four);
        };
        \end{feynman} 
    \end{tikzpicture}
   \caption{1-loop neutrino mass generation of the scotogenic model~\cite{Ma:2006km}.}
   \label{fig:scoto}
\end{figure}
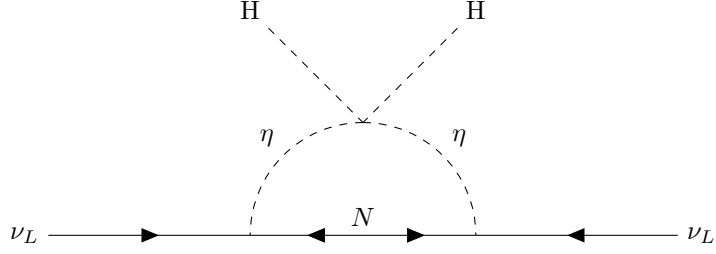
\begin{figure}[!t]
    \centering
    \includegraphics[width=0.8\linewidth]{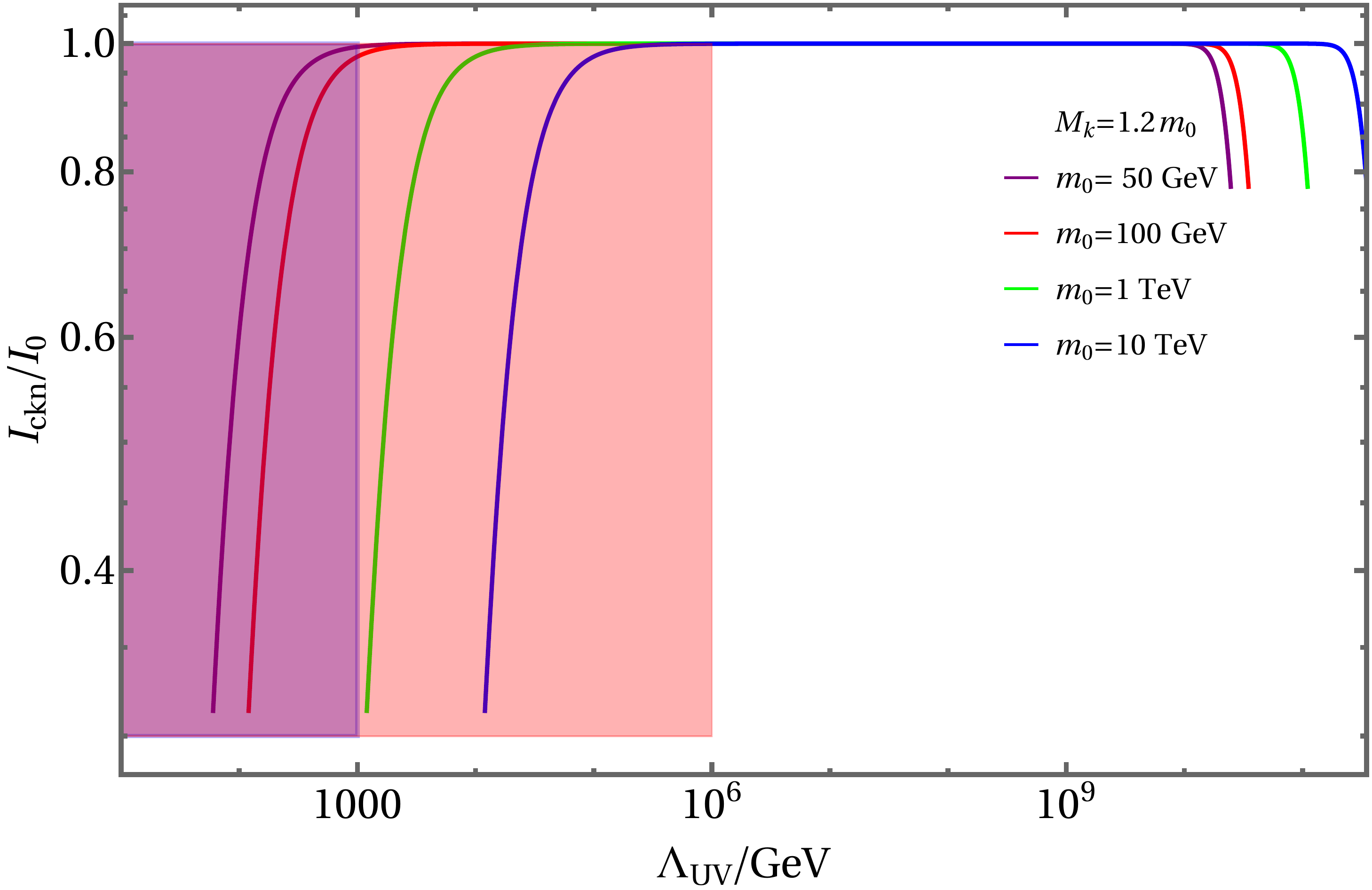}
    \caption{Relative discrepancy between the neutrino mass with and without the influence of the CKN bound for the scotogenic model. Different mass choices for the free parameters are considered. The allowed range of $\Lambda_{UV}$ in the calculation of the magnetic moment of the muon is displayed as a red background. The area allowed by both the magnetic moment of the electron and muon is shown as the violet region.}
    \label{fig:scoto_ckn}
\end{figure}
For the calculation of the neutrino mass matrix we split the neutral scalar component in its real and imaginary part $\eta^{0}=\frac{1}{\sqrt{2}}(\eta_{R}+i\eta_{I})$ and only the neutral component of the Higgs doublet $H$ receives a vacuum expectation value of $v$. We obtain the expression 
\begin{equation}\label{eq:scoto}
    (M_{\nu})_{ij}=\sum_{k} \frac{h_{ik} h_{jk}M_k}{32\pi^2} \left(\frac{m_R^2}{m_R^2-M_k^2}\ln\left(\frac{m_R^2}{M_k^2}\right)-\frac{m_I^2}{m_I^2-M_k^2}\ln\left(\frac{m_I^2}{M_k^2}\right)\right)\text,
\end{equation}
where the $M_k$ are the masses of the singlet fermions $N_k$, $m_{R/I}$ denote the masses of the real/imaginary part of the neutral scalar component and $h_{ik}$ are the Yukawa couplings (note that the missing factor of 1/2 in the original work~\cite{Ma:2006km} has been corrected, in agreement with Ref.~\cite{Cai:2017jrq}). 
After considering the limit $m_R^2-m_I^2=2\lambda_5 v^2\ll m_0^2$, with $m_0^2=\frac{m_R^2+m_I^2}{2}$, the result simplifies to 
\begin{align}
    (M_{\nu})_{ij}&\approx \frac{\lambda_5 v^2}{16\pi^2}\sum_{k} h_{ik} h_{jk} I_{0,scoto}(M_k,m_0)
\end{align}
with 
\begin{align}\label{eq:scoto_lit}
    I_{0,scoto}(M_k,m_0)=\frac{M_k}{m_0^2-M_k^2}\left(1-\frac{M_k^2}{m_0^2-M_k^2}\ln\left(\frac{m_0^2}{M_k^2}\right)\right)\text,
\end{align}
which reduces the number of free parameters by one.

If we consider the CKN bound and introduce momentum cutoffs in our calculation, Eq. \eqref{eq:scoto} changes to 
\begin{equation}
\begin{aligned}
    (M_{\nu})_{ij,ckn}&=\sum_{k}\frac{h_{ik}h_{jk}M_k}{32\pi^2}\Biggl(\frac{m_R^2}{-m_R^2-M_k^2}\ln\left(\frac{(\Lambda_{UV}^2+m_R^2)(\Lambda_{IR}^2+M_k^2)}{(\Lambda_{UV}^2+M_k^2)(\Lambda_{IR}^2+m_{R}^2)}\right)
    \\&+\frac{m_I^2}{m_I^2-M_k^2}\ln\left(\frac{(\Lambda_{UV}^2+m_I^2)(\Lambda_{IR}^2+M_k^2)}{(\Lambda_{UV}^2+M_k^2)(\Lambda_{IR}^2+m_{I}^2)}\right)\Biggr) 
\end{aligned}
\end{equation}
and with the limit $2 \lambda_5 v^2\ll m_0^2$ we get
\begin{align}
    (M_{\nu})_{ij,ckn}\approx\frac{\lambda_5 v^2}{16\pi^2}\sum_{k} h_{ik} h_{jk} I_{ckn,scoto}(M_k,m_0,\Lambda_{UV},\Lambda_{IR}(\Lambda_{UV}))
\end{align}
with 
\begin{align}\label{eq:scoto_ckn}
    &I_{ckn,scoto}(M_k,m_0,\Lambda_{UV},\Lambda_{IR}(\Lambda_{UV}))=
    \frac{M_k}{m_0^2-M_k^2}\nonumber\\&\times \Biggl(-\frac{m_0^2(\Lambda_{IR}^2-\Lambda_{UV}^2)}{(m_0^2+\Lambda_{IR}^2)(m_0^2+\Lambda_{UV}^2)}
    +\frac{M_k^2}{m_0^2-M_k^2}\ln\left(\frac{(m_0^2+\Lambda_{UV}^2)(M_k^2+\Lambda_{IR}^2)}{(m_0^2+\Lambda_{IR}^2)(M_k^2+\Lambda_{UV}^2)}\right)\Biggr) \text.
\end{align}

In figure \ref{fig:scoto_ckn}, we plot the ratio $I_{ckn,scoto}/I_{0,scoto}$ of the simplified integrals with and without the CKN bound (Eq. \eqref{eq:scoto_lit} and \eqref{eq:scoto_ckn}) in dependence of the UV cutoff $\Lambda_{UV}$. For clarity we consider only the impact on one non-vanishing neutrino mass. In this case one has to fix one singlet fermion mass $M_k$ and the Yukawa coupling cancel out in the ratio considered. The remaining free parameters are $m_0$ and $M_k$ that are chosen as follows. Since we are mainly interested in the impact of the CKN bound on the model predictions and do not aspire to make precise predictions, we fix $M_k=1.2\cdot m_0$ so that $\eta_R$ or $\eta_I$ is a possible dark matter candidate with mass $m_0$, for which we adopt values which approximately cover the possible dark matter mass range (\(m_{0} = 50\;\mathrm{GeV},100\;\mathrm{GeV}, 1\;\mathrm{TeV},10\;\mathrm{TeV}\)).
The result is plotted in figure \ref{fig:scoto_ckn}, where we also assumed that the masses running in the loop diagrams considered are confined to the interval $[\Lambda_{IR};\Lambda_{UV}]$. Accordingly, we obtain different start and endpoints of the plots corresponding to different choices of masses $m_0$.
In the case of a universal cutoff, which has to be inside the overlap of the red and blue areas, we obtain bounds for the allowed masses $m_0$. In the cases considered, just the masses \(m_{0} = 50\;\mathrm{GeV},100\;\mathrm{GeV}\) would be possible (dependent on the proportionality factor between the masses $m_0$ and $M_k$) and the CKN bound can reduce the integral and thus the neutrino masses by up to $\sim 80 \%$. If, on the other hand, the choice of the cutoff depends on the process, the comparison with the magnetic moment would not imply a constraint and the amount of suppression would depend on the concrete value of $\Lambda_{UV}$. 

\subsection{Zee model (T-I-2)}
\begin{figure}[!t]
    \centering
    \begin{tikzpicture}[x=3cm, y=3cm]
        \begin{feynman}
            \vertex (a) at (0,0) {\( \nu_L \)};
            \vertex (b) at (1,0); 
            \vertex (bc) at (1.5,0); 
            \vertex (c) at (2,0);
            \vertex (d) at (3,0){\( \nu_L \)};
            \vertex (h1) at (1.5,1){H};
            \vertex (h2) at (1.5,-0.5){H};
            \vertex (four) at (1.5,0.5);
            \diagram* {
            (a)--[fermion](b);
            (d)--[fermion] (c);
            (bc)--[fermion, edge label'=\(e_L\)] (b);
            (bc)--[fermion, edge label=\(e_R\)] (c);
            (four)--[scalar, quarter right, edge label'=\(s^{+}\)](b);
            (four)--[scalar, quarter left,edge label=\(\Phi^{+}\)](c);
            (h1)--[scalar](four);
            (h2)--[scalar](bc);
            };
            \end{feynman} 
    \end{tikzpicture}
   \caption{1-loop neutrino mass generation of the Zee model~\cite{Zee:1980ai}.}
   \label{fig:zee}
\end{figure}
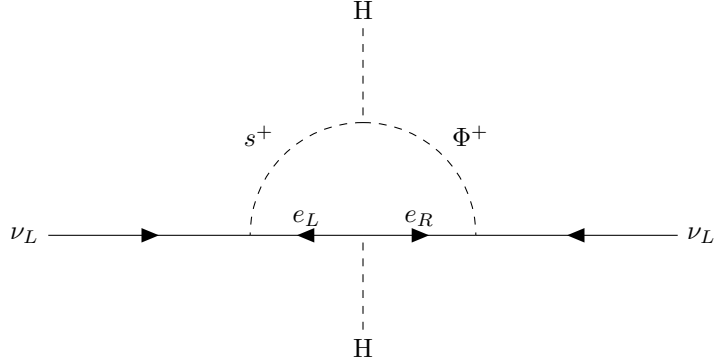
The Zee model~\cite{Zee:1980ai} extends the SM with an additional Higgs scalar doublet $\Phi$ and a positive charged scalar singlet $s$.

The relevant part of the Lagrangian is given by 
\begin{equation}
    -\mathcal{L}_{Z}^{Y}=\bar{L}\left(y_{e} H+\Gamma_{e} \Phi\right) e_{R}+y_{s} s \overline{L^{c}} L+ h.c.  \text,
\end{equation}
where the Yukawa couplings $y_e$ and $\Gamma_e$ are given by in general complex $3 \times 3$ matrices and the coupling $y_s$ in the second term has to be antisymmetric (see e.g. \cite{Cordero-Carrion:2019qtu,Cai:2017jrq}). To close the loop, we also need the term
\begin{equation}
    \mu_{Z} H \Phi s^* + h.c.
\end{equation}
of the scalar potential, with $\mu_Z$ being a parameter of dimension mass. We also consider that only the Higgs scalar gets an vacuum expectation value $v$. 
Mixing between the flavor states of the new particles ($\Phi^+,s^+$) occurs, where
\begin{equation}
    \sin(2\phi)=\frac{\sqrt{2}v\mu_{Z}}{M_{2}^2-M_{1}^2}
\end{equation}
is the mixing angle and $M_1,M_2$ are the masses of the charged particles. We chose a basis in which the Yukawa matrix of the charged leptons 
\begin{equation}
    M_l=\frac{y_l v}{\sqrt{2}}
\end{equation}
is diagonal with the mass eigenvalues being $m_{l_i}$.
\begin{figure}[!t]
    \centering
    \includegraphics[width=0.8\linewidth]{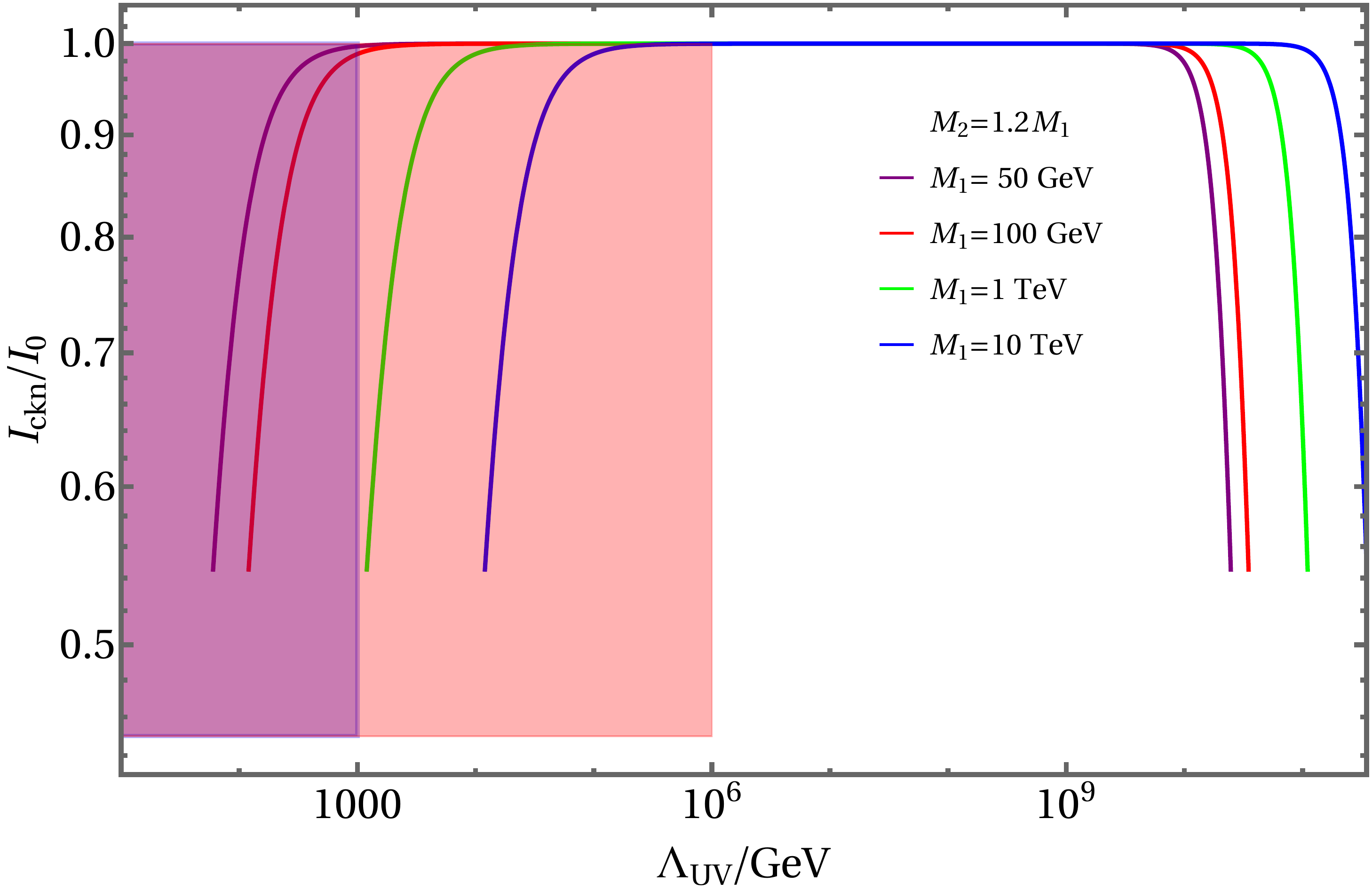}
    \caption{As figure \ref{fig:scoto_ckn} for the Zee model.}
    \label{fig:zee_ckn}
\end{figure}
It follows that the lowest order contribution to the neutrino mass is given by the diagram shown in figure \ref{fig:zee} and leads to the neutrino mass matrix 
\begin{equation}
    M_\nu=\sum_i -\frac{\sin(2\phi)}{16\pi^2}(y_s M_l \Gamma_e + \Gamma_e^T M_l y_s^T)I_{0,zee}(M_1,M_2,m_{l_i})\text,
\end{equation}
with
\begin{equation}
    I_{0,zee}(M_1,M_2,m_{l_i})=\left(\frac{M_2^2}{M_2^2-m_{l_i}^2}\log\left(\frac{M_2^2}{m_{l_i}^2}\right)-\frac{M_1^2}{M_1^2-m_{l_i}^2}\log\left(\frac{M_1^2}{m_{l_i}^2}\right)\right)\text.
\end{equation}
On adopting the CKN bound, $I_{0,zee}(M_1,M_2,m_{l_i})$ changes to 
\begin{equation}
\begin{aligned}
    &I_{ckn,zee}(M_1,M_2,m_{l_i},\Lambda_{UV},\Lambda_{IR}(\Lambda_{UV}))= \\
    &\left(\frac{-M_2^2}{M_2^2-m_{l_i}^2}\log\left(\frac{(\Lambda_{UV}^2+M_2^2)(\Lambda_{IR}^2+m_{l_i}^2)}{(\Lambda_{UV}^2+m_{l_i}^2)(\Lambda_{IR}^2+M_{2}^2)}\right)+\frac{M_1^2}{M_1^2-m_{l_i}^2}\log\left(\frac{(\Lambda_{UV}^2+M_1^2)(\Lambda_{IR}^2+m_{l_i}^2)}{(\Lambda_{UV}^2+m_{l_i}^2)(\Lambda_{IR}^2+M_{1}^2)}\right)\right) \text.
\end{aligned}
\end{equation}

By assuming that the new particles are heavier than the SM leptons, $m_{l_i}^2\ll M_1^2,M_2^2$, we can approximate the results with 
\begin{equation}
    I_{0,zee}(M_1,M_2,m_{l_i})\approx \log\left(\frac{M_2^2}{M_1^2}\right)
\end{equation}
and obtain
\begin{equation}
    I_{ckn,zee}(M_1,M_2,m_{l_i},\Lambda_{UV},\Lambda_{IR}(\Lambda_{UV}))\approx \log\left(\frac{(\Lambda_{UV}^2+M_1^2)(\Lambda_{IR}^2+M_2^2)}{(\Lambda_{IR}^2+M_1^2)(\Lambda_{UV}^2+M_2^2)}\right) \text.
\end{equation}

As before with the previous model, we show the relative deviation between the results with and without CKN bound in figure \ref{fig:zee_ckn}, where we also consider $M_2=1.2\cdot M_1$. The effect is similar to the scotogenic model, but now the effect in the high energy range is stronger and the maximum effect is reduced to approximately $50 \%$.

\subsection{Inverse scotogenic model (T-I-3)}
\begin{figure}[!t]
    \centering
    \begin{tikzpicture}[x=3cm, y=3cm]
        \begin{feynman}
            \vertex (a) at (0,0) {\( \nu_L \)};
            \vertex (b) at (1,0); 
            \vertex (bc) at (1.5,0); 
            \vertex (c) at (2,0);
            \vertex (d) at (3,0){\( \nu_L \)};
            \vertex (h1) at (1,1){H};
            \vertex (h2) at (2,1){H};
            \vertex (four2) at (1.75,0.433);
            \vertex (four1) at (1.25,0.433);
            \diagram* {
            (a)--[fermion](b);
            (d)--[fermion] (c);
            (b)--[fermion, bend left, edge label=\(E^{0}_R\)](four1);
            (four1)--[majorana,bend left, edge label=\(N_L\)](four2);
            (c)--[fermion, bend right,edge label'=\(E^{0}_R\)](four2);
            (h1)--[scalar](four1);
            (h2)--[scalar](four2);
            (b)--[scalar,edge label=s] (c)
            };
            \end{feynman} 
    \end{tikzpicture}
   \caption{1-loop neutrino mass generation of the inverse scotogenic model~\cite{Fraser:2014yha,Ma:2015pma}.}
   \label{fig:invScoto}
\end{figure}
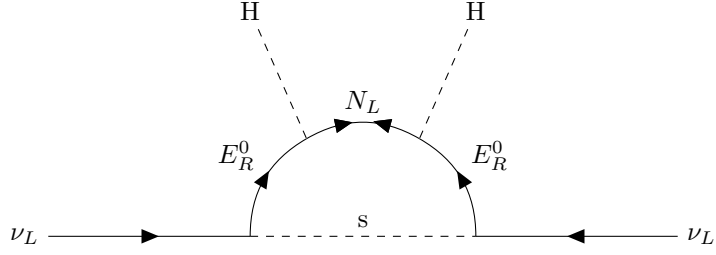      
As a third model we investigate the inverse scotogenic model~\cite{Fraser:2014yha,Ma:2015pma}, which extends the SM by three real singlet scalars $s_i$, one fermion doublet $(E^0,E^-)_{L,R}$ and one singlet fermion $N_L$ with an additional Majorana mass (note that in Ref. \cite{Ma:2015pma} also $N_R$ was considered). In this model, the lightest of the three singlet fermions $s_i$ is the dark matter candidate.
The Lagrangian of this model is given by 
\begin{equation}
    \begin{aligned} \mathcal{L} & =-M_{E}\left(\bar{E}^{0} E^{0}+\bar{E}^{-} E^{-}\right)-\frac{1}{2} M_{N} N_{L} N_{L}+\frac{1}{2}\left(M_{S}^{2}\right)_{i j} s_{i} s_{j} \\ & +f_{d} \bar{N}_{L}\left(E_{R}^{0} \phi^{0}-E_{R}^{-} \phi^{+}\right)+f s_{i}\left(\bar{E}_{R}^{0} \nu_{i L}+\bar{E}_{R}^{-} l_{i L}\right)+h . c .\text,\end{aligned}
\end{equation}
which leads to the mass matrix $\left(\bar{E}_{R}^{0}, E_{L}^{0}, N_{L}\right)$
\begin{equation}
    M_{E, N}=\left(\begin{array}{ccc}0 & M_{E} & M_{D} \\ M_{E} & 0 & 0 \\ M_{D} & 0 & M_{N}\end{array}\right) \text.
\end{equation}
For the calculation we use the limit $M_N \rightarrow 0$ and just keep the terms of linear order in $M_N$. We get, as a result of the one-loop mass generation shown in figure \ref{fig:invScoto}, for the one-loop neutrino mass
\begin{equation}
    m_\nu=\frac{f^2 M_D^2  M_N}{16\pi^2\left(M_D^2 + M_E^2 - M_S^2\right)^2} I_{0,invscoto}(M_D,M_E,M_S) \text,
\end{equation}
with 
\begin{equation}
    I_{0,invscoto}(M_D,M_E,M_S) = \left(M_D^2 + M_E^2 - M_S^2 - M_S^2 \log\left(\frac{M_D^2 + M_E^2}{ M_S^2}\right)\right) \text.
\end{equation}

This changes with the CKN bound to 
\begin{equation}
\begin{aligned}
    &I_{0,invscoto}(M_D,M_E,M_S)\rightarrow I_{ckn,invscoto}(M_D,M_E,M_S,\Lambda_{UV},\Lambda_{IR}(\Lambda_{UV}))= \\
    &\Biggl(M_S^2 \left(\log \left(\frac{\Lambda_{UV}^2+M_D^2+M_E^2}{\Lambda_{UV}^2+M_S^2}\right)-\log \left(\frac{\Lambda_{IR}^2+M_D^2+M_E^2}{\Lambda_{IR}^2+M_S^2}\right)\right)\\
    &-\frac{(\Lambda_{IR}^2-\Lambda_{UV}^2) \left(M_D^2+M_E^2\right) \left(M_D^2+M_E^2-M_S^2\right)}{\left(\Lambda_{IR}^2+M_D^2+M_E^2\right) \left(\Lambda_{UV}^2+M_D^2+M_E^2\right)}\Biggr) \text.
\end{aligned}
\end{equation}

\begin{figure}[!t]
    \centering
    \includegraphics[width=0.8\linewidth]{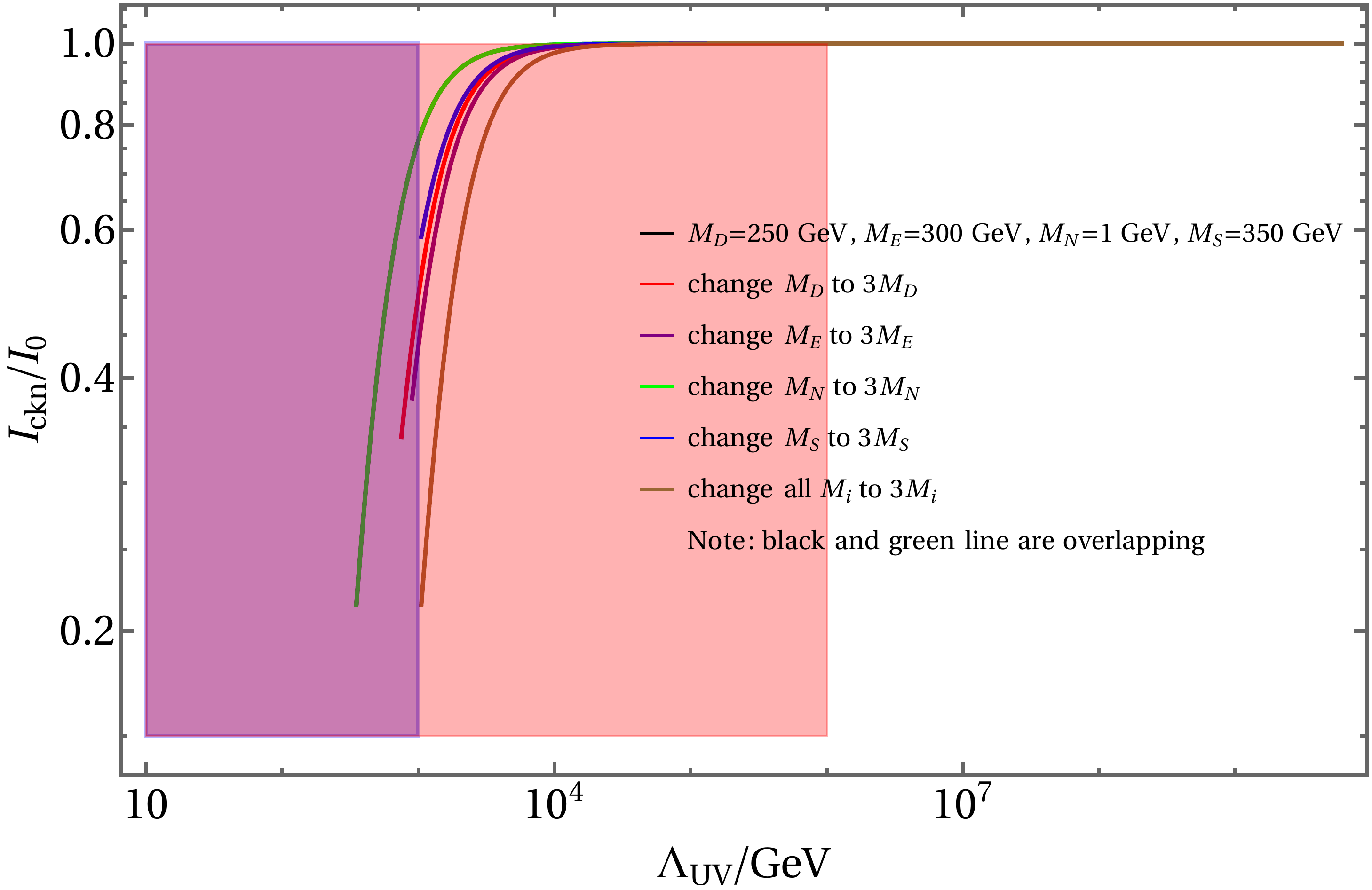}
    \caption{As figure \ref{fig:scoto_ckn} for the inverse scotogenic model.}
    \label{fig:invScoto_ckn}
\end{figure}

This model features several free parameters. To represent the result graphically, we set the masses to \(M_D = 250\;\mathrm{GeV}, M_E=300\;\mathrm{GeV}, M_N= 1 \;\mathrm{GeV}, M_S=350\;\mathrm{GeV}\) and study the effect of the variation of the mass parameters in figure \ref{fig:invScoto_ckn}.
We notice that not all parameters have a strong impact on the result and, different from the other models, the variation does not simply shift the curve to lower or higher energies but also affects the maximum deviation significantly. With the set of parameters considered we observe that the effect results to an suppression of up to 40 \% - 80 \% that is constraint to the low energy regime of the curve.

\subsection{ScotoSinglet model (T-I-1)}
\begin{figure}[H]
    \centering
    \begin{tikzpicture}[x=3cm, y=3cm]
        \begin{feynman}
            \vertex (a) at (0,0) {\( \nu_L \)};
            \vertex (b) at (1,0); 
            \vertex (bc) at (1.5,0); 
            \vertex (c) at (2,0);
            \vertex (d) at (3,0){\( \nu_L \)};
            \vertex (h1) at (1,1){H};
            \vertex (h2) at (2,1){H};
            \vertex (four2) at (1.75,0.433);
            \vertex (four1) at (1.25,0.433);
            \diagram* {
            (a)--[fermion](b);
            (d)--[fermion] (c);
            (b)--[scalar, bend left, edge label=\(\Phi\)](four1);
            (four1)--[scalar,bend left, edge label=\(\varphi\)](four2);
            (c)--[scalar, bend right,edge label'=\(\Phi\)](four2);
            (h1)--[scalar](four1);
            (h2)--[scalar](four2);
            (b)--[majorana,edge label=\(\Psi\)] (c)
            };
            \end{feynman} 
    \end{tikzpicture}
   \caption{1-loop neutrino mass generation diagram of the ScotoSinglet model~\cite{Beniwal:2020hjc}, which contributes additionally to the diagram of the scotogenic model in figure \ref{fig:scoto} with accordingly renamed fields.}
   \label{fig:scotoSinglet}
\end{figure}
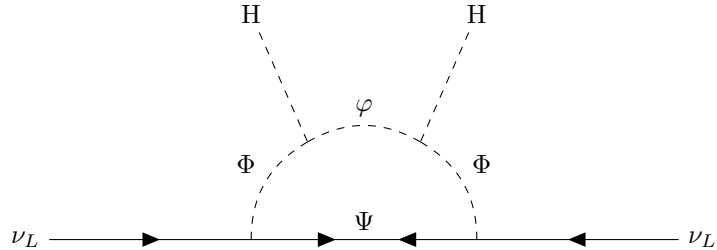
\begin{figure}[!t]
    \centering
    \includegraphics[width=0.8\linewidth]{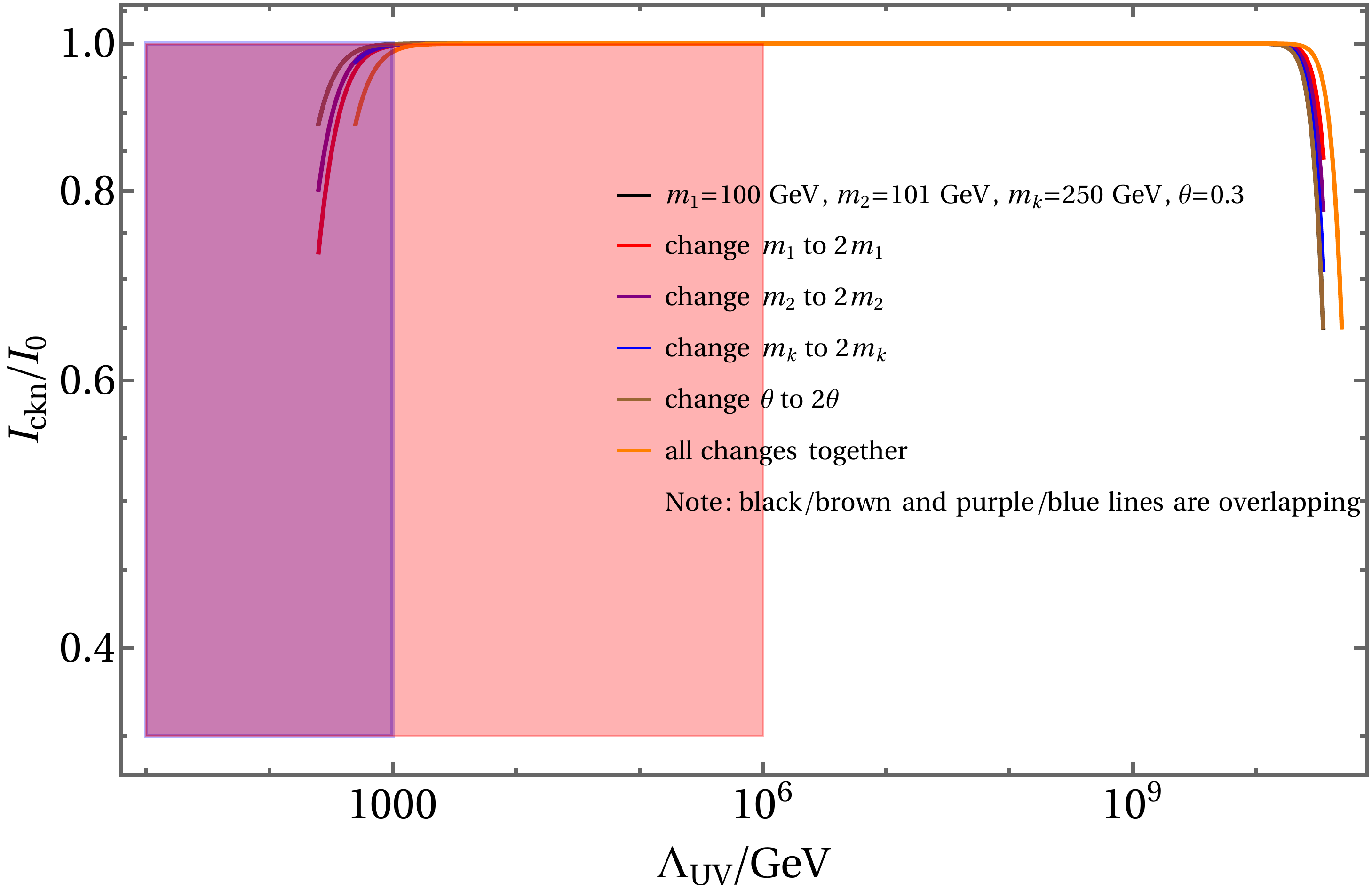}
    \caption{As figure \ref{fig:scoto_ckn} for the ScotoSinglet model.}
    \label{fig:scotoSinglet_ckn}
\end{figure}

The last model we consider is the ScotoSinglet model~\cite{Beniwal:2020hjc} which extends the scotogenic model by a new real scalar singlet $\varphi$ and, together with a $\mathbb{Z}_2$ symmetry, leads to the potential 
\begin{equation}
    \begin{aligned} V= & -\mu_{H}^{2} H^{\dagger} H+\lambda_{H}\left(H^{\dagger} H\right)^{2}+m_{\Phi}^{2} \Phi^{\dagger} \Phi+\lambda_{\Phi}\left(\Phi^{\dagger} \Phi\right)^{2}+\frac{1}{2} m_{\varphi}^{2} \varphi^{2}+\frac{1}{4} \lambda_{\varphi} \varphi^{4} \\ & +\lambda_{H \Phi, 1}\left(H^{\dagger} H\right)\left(\Phi^{\dagger} \Phi\right)+\lambda_{H \Phi, 2}\left(H^{\dagger} \Phi\right)\left(\Phi^{\dagger} H\right)+\frac{1}{2}\left[\lambda_{H \Phi, 3}\left(H^{\dagger} \Phi\right)^{2}+\text { h.c. }\right] \\ & +\frac{1}{2} \lambda_{H \varphi} H^{\dagger} H \varphi^{2}+\frac{1}{2} \lambda_{\Phi \varphi} \Phi^{\dagger} \Phi \varphi^{2}+\left[\kappa \Phi^{\dagger} H \varphi+\text { h.c. }\right]\text,\end{aligned}
\end{equation} 
where we now adopt as convention for the SM Higgs $H=(0,(v+h)/\sqrt{2})^T$, the new doublet $\Phi=(\Phi^+, (\Phi_R+i A)/\sqrt{2})$, the Majorana fermions $\Psi_k$ and Yukawa couplings $y_{ij}$ to be consistent with Ref. \cite{Beniwal:2020hjc}. 
It follows, that in addition to the diagram already shown in figure \ref{fig:scoto}, the diagram in figure \ref{fig:scotoSinglet} contributes. We see that the real part $\Phi_R$ mixes with the new scalar $\varphi$ to produce the mass eigenstates $m_1$ and $m_2$, parametrized by the angle $\theta$. The masses $m_1$ and $m_2$, together with the masses of the Majorana fermions $m_k$, the mass of the imaginary part $m_A$, the mixing angle $\theta$ and 6 couplings constants are the free parameters of this model. Other parameters such as
\begin{equation}\label{eq:param}
    \lambda_{H \Phi, 3}=\frac{1}{v^{2}}\left(m_{1}^{2} \cos ^{2} \theta+m_{2}^{2} \sin ^{2} \theta-m_{A}^{2}\right)
\end{equation}
can be expressed by them, as discussed in detail in Ref. \cite{Beniwal:2020hjc}. The resulting neutrino mass matrix is
\begin{equation}
    M_{ij}=\sum_{k} \frac{y_{ki} m_k y_{kj}}{32\pi^2} I_{0,scotosinglet}(m_A, m_1, m_2, m_k, \Theta) \text,
\end{equation}
with
\begin{align}
    I_{0,scotosinglet}(m_A, m_1, m_2, m_k, \theta)=&\Biggl(\cos^2(\theta)\frac{m_1^2}{m_1^2-m_k^2}\log\left(\frac{m_1^2}{m_k^2}\right)+\sin^2(\theta)\frac{m_2^2}{m_2^2-m_k^2}\log\left(\frac{m_2^2}{m_k^2}\right)\nonumber\\
    &-\frac{m_A^2}{m_A^2-m_k^2}\log\left(\frac{m_A^2}{m_k^2}\right)\Biggr)\text.
\end{align}
On adopting the CKN bound $I_{0,scotosinglet}(m_A, m_1, m_2, m_k, \theta)$ changes to 
\begin{align}
    &I_{ckn,scotosinglet}(m_A, m_1, m_2, m_k, \theta, \Lambda_{UV},\Lambda_{IR}(\Lambda_{UV}))=\nonumber\\
    & \Biggl(\cos^2(\theta)\frac{m_1^2}{m_1^2-m_k^2}\log\left(\frac{(m_1^2+\Lambda_{IR}^2)(m_k^2+\Lambda_{UV}^2)}{(m_k^2+\Lambda_{IR}^2)(m_1^2+\Lambda_{UV}^2)}\right)
    +\sin^2(\theta)\frac{m_2^2}{m_2^2-m_k^2}\log\left(\frac{(m_2^2+\Lambda_{IR}^2)(m_k^2+\Lambda_{UV}^2)}{(m_k^2+\Lambda_{IR}^2)(m_2^2+\Lambda_{UV}^2)}\right)\nonumber\\
    &-\frac{m_A^2}{m_A^2-m_k^2}\log\left(\frac{(m_A^2+\Lambda_{IR}^2)(m_k^2+\Lambda_{UV}^2)}{(m_k^2+\Lambda_{IR}^2)(m_A^2+\Lambda_{UV}^2)}\right)\Biggr) \text.
\end{align}
To study all four possible topologies individually, we set the quartic coupling to $0$, so that only the diagram in figure \ref{fig:scotoSinglet} contributes to the mass. As is obvious from Eq. \eqref{eq:param}, this results in a number of free parameters reduced by one, and fixes the mass $m_A$ as 
\begin{equation}
    m_{A}^{2}=m_{1}^{2} \cos ^{2} \theta+m_{2}^{2} \sin ^{2} \theta \text.
\end{equation}
As before we plot the deviation due to the CKN cutoff in figure \ref{fig:scotoSinglet_ckn}.

\section{Conclusions}
Usually neglected influences of gravity can have significant effects on radiative corrections in quantum field theories. In this paper we studied 
the impact of the CKN bound on the phenomenology of radiative neutrino mas models. We found that the cutoffs imposed by the CKN bound entail dramatic consequences for the parameter space of the models studied. This implies that statements about the possibility to probe the models for example in current experiments, the validity of proposed particles as dark matter candidates, or the question when models can be considered to be ruled out by experimental data have to be reconsidered in the light of the effects studied here. 

We have focused on four exemplary models representing the four different possible topologies for Feynman diagrams generating neutrino masses on 1-loop order. We have shown that the CKN bound can have a significant impact on the masses generated, dependent on the actual choice of the UV cutoff. This effect seems to be independent of the topology of the diagram, and depends mainly on the choice of parameters. The behavior of the different results is always similar, having a strong impact on the lower and higher energy range and very low impact in the middle. If we consider a UV cutoff compatible with the measurements of the magnetic moment of the muon/electron, it is possible to derive constraints on the possible masses of the new particles.

\section*{Acknowledgements}
P.A. thanks the IFIC for hospitality and acknowledges support of the Erasmus+ programme of the European Union during his stay there.
M.H. acknowledges support by grants PID2020-113775GB-I00 (AEI/10.13039/501100011033) and CIPROM/2021/054 (Generalitat Valenciana).


\end{document}